\documentclass[twocolumn,floatfix,superscriptaddress,longbibliography,PRL]{revtex4}
\usepackage{amsfonts,amssymb,amsmath,hyperref}
\usepackage[applemac]{inputenc}
\usepackage{color}
\usepackage{dcolumn}
\usepackage{slashed}
\usepackage{natbib}
\usepackage{diagbox}
\usepackage{bm}

\newif\ifhyper
\hypertrue
\ifhyper
\hypersetup{
citecolor = {green},
colorlinks = {true}, 
urlcolor = {blue} 
}
\fi

\newlength{\ldag}
\def\eps{\epsilon}

\def\be{\begin{equation}}
\def\ee{\end{equation}}
\def\bea{\begin{eqnarray}}
\def\eea{\end{eqnarray}}
\def\bse{\begin{subequations}}
\def\ese{\end{subequations}}
\def\bc{\begin{center}}
\def\ec{\end{center}}

\def\nonum{\nonumber}

\newcommand{\vcrule}{\rule[-0.4cm]{0cm}{1cm}}

\newcommand{\smallvcrule}{\rule[-0.2cm]{0cm}{0.6cm}}
\allowdisplaybreaks

\begin{document}

\title{Three-loop 
 order
approach to flat polymerized membranes}

\author{S. Metayer} 
\email{smetayer@lpthe.jussieu.fr}
\affiliation{Sorbonne Universit\'e, CNRS, Laboratoire de Physique Th\'eorique et Hautes Energies, LPTHE, 75005 Paris, France}

\author{D. Mouhanna} 
\email{mouhanna@lptmc.jussieu.fr}
\affiliation{Sorbonne Universit\'e, CNRS, Laboratoire de Physique Th\'eorique de la Mati\`ere Condens\'e, 75005 Paris, France}

\author{S. Teber} 
\email{teber@lpthe.jussieu.fr}
\affiliation{Sorbonne Universit\'e, CNRS, Laboratoire de Physique Th\'eorique et Hautes Energies, LPTHE, 75005 Paris, France}


\begin{abstract}

We derive the three-loop order renormalization group equations that describe the flat phase of polymerized membranes within the modified minimal subtraction scheme, following the pioneering one-loop order computation of Aronovitz and Lubensky [Phys.\ Rev.\ Lett.\ {\bf 60}, 2634 (1988)] and the recent two-loop order one of Coquand, Mouhanna and Teber [Phys.\ Rev.\ E {\bf 101}, 062104 (2020)]. We analyze the fixed points of these equations and compute the associated field anomalous dimension $\eta$ at three-loop order. Our results display a marked  proximity with those obtained using nonperturbative techniques and reexpanded in powers of $\epsilon=4-D$. Moreover, the three-loop order value that we get for $\eta$ at the stable fixed point, $\eta=0.8872$, in $D=2$,  is compatible with known theoretical results and within  the range of accepted numerical values.

\end{abstract} 

\maketitle 

{\it Introduction.} The flat phase of polymerized membranes has  been recently  the subject of intense investigations mainly motivated by the fact that it seems to encode in a satisfying way the elastic degrees of freedom of materials such as graphene \cite{novoselov04,novoselov05} and, more generally, graphene-like systems
(see, e.g., \cite{katsnelson12}). Early, first order, perturbative computations \cite{nelson87,aronovitz88} have revealed the stability of such a phase, ensured by a mechanism of coupling of the capillary (flexural),  ${\bf h}$, modes with the elastic (phonons),  ${\bf u}$,  modes, allowing to circumvent the Mermin-Wagner theorem; see, e.g. \cite{coquand19b} for an explanation. This flat phase is controlled by a fully stable infrared fixed point, characterized by power-law behaviors for the phonon-phonon and flexural-flexural correlation functions \cite{aronovitz88,guitter88,aronovitz89,guitter89}:
\begin{equation}
G_{uu}(q)\sim q^{-(2+\eta_u)} \hspace{0.7cm} {\hbox{and}} \hspace{0.7cm} G_{hh}(q)\sim {q^{-(4-\eta)}} 
\label{correlation}
\end{equation}
where $\eta_u$ and $\eta$ are nontrivial anomalous dimensions related by a Ward identity: $\eta_u=4-D-2\eta$ \cite{aronovitz88,guitter88,aronovitz89,guitter89}. A major challenge, in this context, is an accurate determination of the exponent $\eta$ at the stable fixed point. Due to the distance between the upper critical dimension,  $D_{\text{uc}}=4$, and the physical dimension, $D=2$, as well as of the intricacy of the diagrammatic analysis involved in the perturbative approach of membranes, the pioneering works have been followed by various nonperturbative approaches able to tackle the physics directly in dimension $D=2$: $1/d$ expansion \cite{david88,guitter88,aronovitz89,guitter89,gornyi15,saykin20}, self-consistent screening approximation (SCSA) \cite{ledoussal92,gazit09,zakharchenko10,roldan11,ledoussal18} and the so-called nonperturbative renormalization group (NPRG) \cite{kownacki09,braghin10,hasselmann11, essafi11,essafi14,coquand16a,coquand18,coquand20}. The two last ones have produced roughly compatible results: $\eta_{\hbox{\scriptsize scsa}}^{\hbox{\scriptsize l}} \simeq 0.821$ \cite{ledoussal92,ledoussal18} at leading order ($\eta_{\hbox{\scriptsize scsa}}^{\hbox{\scriptsize nl}} \simeq 0.789$ at controversial next-to-leading order \cite{gazit09}) and $\eta_{\hbox{\scriptsize nprg}}=0.849$ \cite{kownacki09}. As for Monte Carlo simulations of membranes, they have also led to scattered values $\eta=0.81(3)$ \cite{zhang93}, $\eta= 0.750(5)$ \cite{bowick96}, and $\eta=0.795(10)$ \cite{troster13} and Monte Carlo simulations of graphene to $\eta\simeq 0.85$ \cite{los09}. In order to get a better understanding of the structure of the underlying field-theory, several groups have, very recently, engaged in perturbative studies of both pure \cite{mauri20,coquand20a} and disordered membranes \cite{coquand20b} going beyond leading order. The two-loop order approach performed in particular in \cite{coquand20a} has revealed an intriguing agreement between the perturbative and nonperturbative approaches in the vicinity of the upper critical dimension. Moreover, the value of the two-loop order anomalous dimension in $D=2$, $\eta^{2l}= 0.9139$ \cite{coquand20a}, when compared to the one-loop order one, $\eta^{1l}=0.96$ \cite{aronovitz88,guitter88,aronovitz89,guitter89}, has been found to move in the right direction when referring to the generally accepted values that lie in the range $[0.72,0.88]$. 

We extend here the work done in \cite{coquand20a} by means of a {\sl three-loop} order, weak-coupling, perturbative approach performed near the upper critical dimension $D_{\text{uc}}=4$ within the modified minimal subtraction ($\overline{\rm MS}$) scheme. We compute the renormalization group (RG) equations at this order for both the  flexuron-phonon  {\sl two-field} model as well as for the flexural {\sl effective} model, which are both defined below. We determine the fixed points and the corresponding field anomalous dimensions at order $\epsilon^3$. We finally compare our results to those obtained within the nonperturbative context either reexpanded in powers of $\epsilon$ or directly in the physical dimensions $D=2$. 

As will be seen in the following, our analysis confirms unambiguously the order-by-order agreement between perturbative and nonperturbative approaches, already identified in our previous work \cite{coquand20a}. Moreover, the value that we get for the three-loop order anomalous dimension in $D=2$, $\eta^{3l}=0.8872$, is  compatible with the analytical and numerical nonperturbative results. Such a fast convergence raises the issue of the unusual nature of the series in $\epsilon$ obtained in this context.


{\it The models.} We now present the two models studied here. One describes a membrane as a $D$-dimensional manifold embedded in a $d$-dimensional Euclidean space. The parametrization of a point ${\bf x}\in \mathbb{R}^D$ in the membrane is realized through the mapping ${\bf x}\to {\bf R} ({\bf x})$ with ${\bf R}\in \mathbb{R}^d$. The flat configuration of a membrane is given by ${\bf R}^0({\bf x)}=({\bf x},{\bf 0}_{d_c})$ where ${\bf 0}_{d_c}$ is the null vector of co-dimension $d_c=d-D$. To parametrize the fluctuations around this configuration one decomposes the field $\bf{ R}$ into $\bf{ R}(\bm{x})=[{\bf x}+{\bf u}({\bf x}), {\bf h}({\bf x})]$ where ${\bf u}$ and ${\bf h}$ represent $D$ longitudinal (phonon)  and $d-D$ transverse (flexural) modes, respectively. 
The action in the flat phase is given by \cite{nelson87,aronovitz88,david88,aronovitz89,guitter88,guitter89,coquand20a}
\begin{equation}
\begin{array}{ll}
S[{\bf h,u}]=\displaystyle \int \text{d}^Dx \hspace{-0.3cm} & \displaystyle \left\{ {\kappa \over 2}\big(\Delta {\bf h} \big)^2 
+ {\lambda\over 2}\, u_{ii}^2 + {\mu}\, u_{ij}^2 \right\}\ ,
\label{action}
\end{array}
\end{equation}
where, as usual, one neglects a term $(\Delta {\bf u})^2$ in the curvature energy contribution $(\Delta {\bf R})^2$. 
In Eq.~(\ref{action}), $u_{ij}$ is the strain tensor that encodes the elastic fluctuations around the flat phase configuration ${\bf R}^0({\bf x)}$:
$u_{ij}={1\over 2}(\partial_{i}{\bf R} . \partial_{j}{\bf R}-\partial_{i}{\bf R}^0 . \partial_{j}{\bf R}^0)={1\over 2}(\partial_{i} {\bf R} . \partial_{j}{\bf R}- \delta_{ij})$. It is given by neglecting nonlinearities in the phonon field ${\bf u}$: 
\begin{equation}
u_{ij}\simeq {1\over 2} \left[\partial_i u_j+\partial_j u_i+ \partial_i {\bf h} . \partial_j {\bf h} \right]\ . 
\label{stress}
\end{equation}
In Eq.~(\ref{action}), $\kappa$ is the bending rigidity constant whereas $\lambda$ and $\mu$ are the Lam\'e (elasticity) coefficients; stability considerations require $\kappa$, $\mu$, and the bulk modulus $B=\lambda+2 \mu/D$ to be all {\sl positive}. The action 
(\ref{action}) together with 
Eq.~(\ref{stress}) defines the two-field model. 

Now one can take advantage of the fact that the phonon field ${\bf u}$ appears quadratically in the action (\ref{action}) to integrate over it exactly. In this way one gets an effective action depending only on the flexural field ${\bf h}$. It reads, in Fourier space \cite{ledoussal92,ledoussal18,coquand20a}: 
\begin{widetext}
\begin{flalign}
S_{\text{eff}}[{\bf h}\,] = \frac{\kappa}{2} \int_{\bf k} \, k^4\, |{\bf h}({\bf k})|^2 + 
+\frac{1}{4}\int_{{\bf k}_1,{\bf k}_2,{\bf k}_3,{\bf k}_4} \hspace{-1.2cm}{\bf h}({\bf k}_1) . {\bf h}({\bf k}_2)\,
R_{ab,cd}({\bf q})\,k_1^a\,k_2^b\,k_3^c\,k_4^d\,\,{\bf h}({\bf k}_3) . {\bf h}({\bf k}_4)\, ,
\label{actioneff}
\end{flalign}
\end{widetext}
where $\int_{\bf k}=\int d^D{ k}/(2\pi)^D$ and ${\bf q} ={\bf k}_1 + {\bf k}_2 = -{\bf k}_3 -{\bf k}_4$. The ${\bf q}$-transverse tensor $R_{ab,cd}({\bf q})$ reads \cite{ledoussal92,ledoussal18,coquand20a}:
\begin{equation}
R_{ab,cd}({\bf q}) = b\ N_{ab,cd}({\bf q}) + \mu\,M_{ab,cd}({\bf q})
\label{R}
\end{equation}
with $N_{ab,cd}$ and $M_{ab,cd}$ given by: 
\begin{equation}
\begin{array}{ll}
&\hspace{-0.4cm} N_{ab,cd}({\bf q}) =\displaystyle {1\over D-1}\,P_{ab}^T({\bf q}) \,P_{cd}^{T}({\bf q})\\ \\
&\hspace{-0.4cm}M_{ab,cd}({\bf q}) =\displaystyle {D-1\over 2}\,\big[ N_{ac,bd}({\bf q}) + N_{ad,bc}({\bf q})\big] - N_{ab,cd}({\bf q})
\label{tensors}
\end{array}
\end{equation}
where $P_{ab}^{T}({\bf q})=\delta_{ab}-q_aq_b/{\bf q}^2$ is the transverse projector. In Eq.~(\ref{R}), we have defined the coupling constant $b= {\mu\,(D\lambda + 2\mu)/(\lambda + 2\mu)} $ which is proportional to the bulk modulus $B$ and has a nontrivial $D$-dependence. The action 
(\ref{actioneff}), together with Eqs.~(\ref{R}) and (\ref{tensors}) define the flexural effective model. Working with this model allows a strong check of the computations performed with the action (\ref{action}). 
However, we would like to emphasize that, in order to have a complete correspondence between the physical quantities computed with both the two-field and the flexural effective models, one has to consider $b$ as a $D$-{\it independent} coupling constant when computing within the modified minimal subtraction ($\overline{\rm MS}$) scheme, see \cite{coquand20a,coquand21c}. Moreover, $b$ is also treated as a coupling constant {\it independent} of $\lambda$ and $\mu$.

{\it The renormalization group equations at three-loop order.}  We have derived the RG equations at three-loop order for the flexural-phonon, two-field, model (\ref{action}) and then for the flexural effective model (\ref{actioneff}), both within the modified minimal subtraction ($\overline{\rm MS}$) scheme and in the massless case. As discussed in \cite{guitter89,coquand20a}, the renormalizability of the models relies on Ward identities following the (partially-broken) rotation invariance in the flat phase \cite{guitter88,guitter89}. Our computations have been performed using techniques of massless Feynman diagram calculations, see, e.g., the review \cite{Kotikov:2018wxe}. Their automation has been implemented using Qgraf \cite{Nogueira:1991ex} for the generation of the diagrams, as well as Mathematica to perform the numerator algebra. LiteRed \cite{Lee:2012cn,Lee:2013mka} has also been used to reduce the loop integrals to a finite set of master integrals. 

In the case of the two-field model, we had to evaluate 32 distinct three-loop diagrams for the flexuron self-energy (note that there were only 5 distinct diagrams at two-loop order and 1 diagram at one-loop order) and 19 diagrams for the three-loop phonon self-energy (there were only 3 distinct diagrams at two-loop order and 1 at one-loop order). In the case of the flexural effective model, we had to evaluate 15 distinct three-loop diagrams for the flexuron self-energy (there were only 3 distinct diagrams at two-loop order and 1 diagram at one-loop order) and 11 diagrams for the effective three-loop polarization (there were only 2 distinct diagrams at two-loop order and 1 at one-loop order). As for the masters, the analytic result of \cite{kazakov85,kazakov84,Kotikov:1995cw} was used in order to compute complicated primitively two-loop master diagrams with a noninteger index on the central line. More details will be given in \cite{coquand21c}.

{\it The two-field model.} For the two-field model one introduces the renormalized fields ${\bf h}_R$ and ${\bf u}_R$ through $\displaystyle {\bf h}=Z^{1/2}\kappa^{-1/2}{\bf h}_R $ and ${\bf u}=Z\kappa^{-1} {\bf u}_R$ and the renormalized coupling constants $\lambda_R$ and $\mu_R$ through: $ \lambda=k^\epsilon Z^{-2}\kappa^2Z_\lambda \lambda_R$ and $\mu=k^\epsilon Z^{-2}\kappa^2Z_\mu \mu_R$ where $k$ is the renormalization momentum scale and $\epsilon=4-D$. Within the $\overline{\rm MS}$ scheme, one moreover introduces the scale $\overline{k}^2=4\pi e^{-\gamma_E}k^2$ where $\gamma_E$ is the Euler constant. One then defines the RG $\beta$-functions $\beta_{\lambda_R}=\partial_t\lambda_R$ and $\beta_{\mu_R}=\partial_t\mu_R$, with $t=\ln \overline{k}$ as well as the field anomalous dimension: 
\begin{equation}
\eta=\beta_{\lambda_R}\, {\partial\ln Z\over \partial \lambda_R}+\beta_{\mu_R}\, {\partial\ln Z\over \partial \mu_R}\ . 
\nonumber
\end{equation} 
where, for simplicity, we have omitted all explicit references to $k$ in our notations of the renormalized coupling constants.

The RG functions are too long to be displayed in the main text; they are given in App.~A. Here we  discuss only the fixed points and the corresponding field anomalous dimensions, whose expressions are explicitly provided in App.~B. Similarly to what happens at one \cite{aronovitz88,guitter88,aronovitz89,guitter89} and two-loop \cite{coquand20a} orders, the three-loop order RG equations display four fixed points (note that, for simplicity, we omit the R indices on the renormalized coupling constants):

\vspace{0.2cm}

(1) The Gaussian fixed point $P_1$ with $\mu^*_1=0,\, \lambda^*_1=0$ and $\eta_1=0$; it is twice unstable. 

\vspace{0.2cm}

(2) The shearless fixed point $P_2$ with $\mu^*_2=0,\, \lambda^*_2={16\pi^2\,\epsilon/d_c}$ and $\eta_2=0$ whose coordinates are the same as those obtained at one and two-loop orders. It lies on the stability line $\mu=0$; it is once unstable.

\vspace{0.2cm}

(3) The fixed point $P_3$ whose coordinates $\mu^*_3,\, \lambda^*_3$ and anomalous dimension $\eta_3$ are given in Table~\ref{tablexpo3} of App.~B. It is once unstable. For this fixed point, whose bulk modulus $B$ vanishes at one-loop order, $B_3^*=\lambda^*_3+2\mu^*_3/(4-\epsilon)$ receives negative contributions of order $\epsilon^2$ at two-loop order and of order $\epsilon^3$ at three-loop order. As observed in \cite{coquand20a}, $P_3$ is thus apparently located out of the stability region of the model (\ref{action}) that requires $B\ge 0$. However, as emphasized in \cite{coquand20a}, this could be an artifact of the perturbative computation -- see below. It is instructive to consider the physical case $d_c=1$ for the anomalous dimension. One gets from Table~\ref{tablexpo3} of App.~B: 
\begin{equation}
\eta_3=0.4762\, \epsilon -0.01776\, \epsilon^2-0.00872\, \epsilon^3 +O(\epsilon^4) \, .
\label{seta3pert}
\end{equation} 
This series shows, up to, and including, the order $\epsilon^3$, a strong decrease of its coefficients. 

\vspace{0.2cm}

(4) The flat phase fixed point $P_4$ whose coordinates $\mu^*_4,\, \lambda^*_4$ and anomalous dimension $\eta_4$ are given in Table~\ref{tablexpo4} of App.~B. It is fully stable and thus controls the asymptotic behavior of the flat phase. Note  that, at one-loop order, this fixed point is located on the line $3\lambda+\mu=0$, which  is stable, in the RG sense,  at one-loop order. In $D$ dimensions, it corresponds to the line $(D+2)\lambda+2\mu=0$. However, at two- \cite{coquand20a} and three-loop orders, the coordinates of the fixed point $P_4$ no longer obey the relation $(D+2)\lambda^*_4+2\mu^*_4=(6-\epsilon)\lambda^*_4+2\mu^*_4=0$. This results in the following nonvanishing anomalous ratio:
\begin{equation}
\delta_1=\frac{\lambda_4^*}{\mu_4^*}+\frac{1}{3} = 0.00889\, \epsilon + 0.02434\, \epsilon^2 +O(\epsilon^3)\,,
\label{anomratio1}
\end{equation}
see discussion below Eq.~(\ref{anomratio2}) for further details.

Finally, considering the physical case $d_c=1$ one gets from Table~\ref{tablexpo4} of App.~B: 
\begin{equation}
\eta_4=0.4800\, \epsilon -0.01152\, \epsilon^2-0.00334\, \epsilon^3 + O(\epsilon^4)\,, 
\label{seta4pert}
\end{equation} 
which  was  referred to simply as $\eta$ in the Introduction.
This series in $\epsilon$ behaves in a way similar to Eq.~(\ref{seta3pert}) with even a stronger decrease of the coefficients than at the fixed point $P_3$. 

{\it The flexural effective model.} As said in the Introduction, we have also considered, as in \cite{coquand20a}, the flexural effective model. This model provides a field theory structurally very different from that provided by the two-field model. From a practical point of view, the task seemed to be less imposing as it  involves only a total of 26 distinct diagrams. However, the computational time remains the same as for the 51 diagrams of the two-field model. From a more conceptual point of view, 
analyzing this model allows one to get valuable  insights about phenomena observed in the context of the two-field model that could correspond to artifacts of the corresponding perturbative approach. Indeed, the two models, although nonperturbatively equivalent, are parametrized by different sets of coupling constants, $(\lambda,\mu)$ and $(b,\mu)$ respectively.

As for the two-field model, one introduces the renormalized field ${\bf h}_R$ through $\displaystyle {\bf h}=Z^{1/2}\kappa^{-1/2}{\bf h}_R $, the renormalized coupling constants $b_R$ and $\mu_R$ through $b=k^\epsilon Z^{-2}\kappa^2Z_b\; b_R$ and $\mu=k^\epsilon Z^{-2}\kappa^2Z_\mu\; \mu_R$. One then defines the RG $\beta$-functions $\beta_{b_R}=\partial_t b_R$ and $\beta_{\mu_R}=\partial_t\mu_R$ as well as the field anomalous dimension 
\begin{equation}
\eta=\beta_{b_R}\, {\partial\ln Z\over \partial b_R}+\beta_{\mu_R}\, {\partial\ln Z\over \partial \mu_R}\ . 
\nonumber
\end{equation}

Again, the RG functions are too long to be given in the main text; they are given in App.~C while the fixed points and the corresponding field anomalous dimensions are given in App.~D. We now discuss the fixed points of these equations. At three-loop order one finds four fixed points (once again, we omit the R indices on the renormalized coupling constants): 

\vspace{0.2cm}

(1) The Gaussian one $P_1$ with $ \mu^*_1=0,\, b^*_1=0$ and $ \, \eta_1=0$, which is twice unstable.

\vspace{0.2cm}

(2) A fixed point $P_2'$ with ${\mu}'^{*}_2=0$ and non trivial values for both ${b}'^*_2$ and $\eta_2'$ see Table~\ref{tablexpoeff2} of App.~D. It is once unstable. The series in $\epsilon$ for $\eta_2'$, in the physical $d_c=1$ case, is given in Table~\ref{tableP2}. Note that this fixed point has no counterpart within the two-field model where $b$, which is proportional to $\mu$, vanishes at $P_2'$. One recalls that the two-loop order correction to this fixed point has first been computed by Mauri and Katsnelson \cite{mauri20}. Considering the physical case $d_c=1$ one gets from Table~\ref{tablexpoeff2} of App.~D: 
\begin{equation}
\eta_2'= 0.4000\, \epsilon -0.00133 \, \epsilon^2 + 0.00138\, \epsilon^3 +  O(\epsilon^4)\, ,
\label{seta2'pert}
\end{equation} 
where the coefficients of the $\epsilon$-expansion are still small but, contrary to what is observed in Eqs.~(\ref{seta3pert}) and  (\ref{seta4pert}), the three-loop order coefficient is now slightly higher than the two-loop order one. This may reveal the asymptotic nature of the expansion. 
That this manifests at $P_2'$ rather than at $P_3$ and $P_4$ seems to be due to the structure of the perturbative series involving denominators that are odd powers of $n+d_c$ with $n=4$ at $P_2'$ while $n=20$ at $P_3$ and $n=24$ at $P_4$, see Tabs.~\ref{tablexpoC30}, \ref{tablexpoC3} and \ref{tablexpoC3bis} in App.~E as well as the discussion below. We do not exclude that, at higher orders, the coefficients of the $\epsilon$-expansions in Eqs.~(\ref{seta3pert}) and (\ref{seta4pert}) may increase as well.
\begin{table}[h!]
\begin{center}
\renewcommand{\arraystretch}{1.3}
\begin{tabular}{cc}
\hline
\hline
Approach & $P_2'$   \\ 
\hline
Three-loop  & $\displaystyle \eta_2'= 0.4000\, \epsilon -0.00133 \, \epsilon^2 + 0.00138\, \epsilon^3$ \\
SCSA        & $\displaystyle \eta_2'= 0.4000\, \epsilon -0.00133 \, \epsilon^2 + 0.00310\, \epsilon^3$ \\  
NPRG        & $\displaystyle \eta_2'= 0.4000\, \epsilon +0.00867 \, \epsilon^2 + 0.00123\, \epsilon^3$ \\ \hline
\hline
\end{tabular}
\end{center}
\vspace{-0.25cm}
\caption{Anomalous $\eta_2'$ at order $\epsilon^3$ at the fixed point $P_2'$ obtained from the three-loop order (this work), SCSA and NPRG approaches.}
\label{tableP2}
\end{table}
%


(3) The infinitely compressible fixed point $P_3$ which is characterized by $b^*_3=0$, thus for which the bulk modulus $B$ vanishes, and by non trivial values for $\mu^*_3$ and $\eta_3$, see Table~\ref{tablexpoeff3} of App.~D. It is once unstable. This fixed point identifies with the fixed point $P_3$ of the two-field model. However, as discussed above, the condition $B_3^*=0$ is violated at both two and three-loop orders for the two-field model. Therefore, the result $B_3^*=0$ obtained within the flexural effective model seems to indicate that this should be, in fact, an artifact of the perturbative ($\epsilon$ expansion)  approach performed on the two-field model and, more precisely, of the handling of $D$-dependent relations as $\lambda+2 \mu/D$ that governs the value taken by the bulk modulus at a given fixed point. Despite this, the series in $\epsilon$ for the anomalous dimension $\eta_3$, see Table~\ref{tablexpoeff3} of App.~D, coincides exactly with that obtained within the two-field model, see Table~\ref{tablexpo3} of App.~B. This quantity, in the physical $d_c=1$ case, is given in Table~\ref{tableP3} and obviously coincides exactly with Eq.~(\ref{seta3pert}).

\begin{table}[h!]
\begin{center}
\renewcommand{\arraystretch}{1.3}
\begin{tabular}{cc}
\hline
\hline
Approach & $P_3$   \\ 
\hline
Three-loop  & $\displaystyle \eta_3= 0.4762\, \epsilon -0.01776\, \epsilon^2-0.00872\, \epsilon^3$ \\ 
SCSA        & $\displaystyle \eta_3= 0.4762\, \epsilon -0.01668\, \epsilon^2-0.00700\, \epsilon^3 $ \\ 
NPRG        & $\displaystyle \eta_3= 0.4762\, \epsilon -0.01349\, \epsilon^2-0.00649\, \epsilon^3$ \\ \hline
\hline
\end{tabular}
\end{center}
\vspace{-0.25cm}
\caption{Anomalous $\eta_3$ at order $\epsilon^3$ at the fixed point $P_3$ obtained from the three-loop order (this work), SCSA and NPRG approaches.}
\label{tableP3}
\end{table}


(4) The fixed point $P_4$, whose coordinates $\mu^*_4, b^*_4$ and anomalous dimension $\eta_4$ are given in Table~\ref{tablexpoeff4} of App.~D. It is fully stable and therefore controls the flat phase. It identifies with the fixed point $P_4$ of the two-field model. At two- and three-loop orders, the coordinates of $P_4$ differ from those obtained from the two-field model, see Table~\ref{tablexpo4} of App.~B. Also, these coordinates do not meet the condition $(D+1)b^*_4-2\mu^*_4=(5-\eps)b^*_4-2\mu^*_4=0$ corresponding to the one-loop order stable line, a result already true at two-loop order \cite{coquand20a}. The corresponding anomalous ratio reads, using $\lambda=2\mu\,(\mu-b)/(b- D\,\mu)$: 
\begin{equation}
\delta_2=\frac{\lambda_4^*}{\mu_4^*}+\frac{1}{3} = 0.00519 \, \epsilon + 0.02122\, \epsilon^2 +O(\epsilon^3)\,.
\label{anomratio2}
\end{equation}
One should notice that this ratio is different from the one found via two-field approach, Eq.~(\ref{anomratio1}), {\it i.e.} $\delta_2\neq\delta_1$, therefore implying  that corrections to  $\lambda/\mu$ are very likely scheme-dependent. This should also be the case for  the Poisson ratio that is given by $\nu=\lambda_4^*/(2\mu_4^*+(D-1)\lambda_4^*)$ since one has  $\nu=-1/3+\delta\nu$ with  $\delta\nu(\delta_1)\neq\delta\nu(\delta_2)$. This  contrasts with the anomalous dimension $\eta_4$, see  Table~\ref{tablexpoeff4}  
 of App.~D, that  coincides exactly with  the one obtained within the two-field model, see Table~\ref{tablexpo4}  of App.~B. The corresponding value in the physical $d_c=1$ case is given in Table~\ref{tableP4} and coincides with Eq.~(\ref{seta4pert}). Finally it is worth noticing that, contrary to the case of the anomalous dimensions at the fixed points, the coefficients of the series giving $\delta_1$ -- Eq.~(\ref{anomratio1}) -- and $\delta_2$ -- Eq.~(\ref{anomratio2}) -- increase with the order of the expansion so that Eqs.~(\ref{anomratio1}) and (\ref{anomratio2}) seem to deserve resummations. Performing a simple symmetric Pad\'e with $\epsilon= 2$ reveals that both $\delta$ are very small ($\sim 10^{-3}$), implying small deviations from the line $3\lambda+\mu=0$, and thus a Poisson ratio close to $-{1/3}$.

\begin{table}[h!]
\begin{center}
\renewcommand{\arraystretch}{1.3}
\begin{tabular}{cc}
\hline
\hline
Approach & $P_4$   \\ 
\hline
Three-loop  & $\displaystyle \eta_4= 0.4800\, \epsilon -0.01152\, \epsilon^2-0.00334\, \epsilon^3$ \\  
SCSA        & $\displaystyle \eta_4= 0.4800\, \epsilon -0.01190\, \epsilon^2-0.00349\, \epsilon^3$ \\  
NPRG        & $\displaystyle \eta_4= 0.4800\, \epsilon -0.00918\, \epsilon^2-0.00333 \, \epsilon^3$ \\ \hline
\hline
\end{tabular}
\end{center}
\vspace{-0.25cm}
\caption{Anomalous $\eta_4$ at order $\epsilon^3$ at the fixed point $P_4$ obtained from the three-loop order (this work), SCSA and NPRG approaches.}
\label{tableP4}
\end{table}

{\it Discussion: Comparison with nonperturbative approaches.} We further discuss our results in particular in comparison with those obtained using alternative methods: the SCSA and the NPRG approaches that are known to have produced numerical results in $D=2$ rather close to the accepted -- although dispersed -- values obtained by means of numerical computations. Also, these methods offer explicit expressions of the various anomalous dimensions as functions of $D$ that can be expanded in powers of $\epsilon$ and compared to those obtained in this work. The anomalous dimensions obtained within the SCSA approach \cite{ledoussal92,ledoussal18}, are given in Table~\ref{tablexpoC3} of App.~E, and those obtained within the NPRG approach \cite{kownacki09}, in Table~\ref{tablexpoC3bis} of App.~E. One first has to note that the basic structure of the series, where denominators are odd powers of $n+d_c$ with $n=4, 20,24$, is recovered within all approaches. Also, as already noted in \cite{coquand20a}, the agreement between the perturbative approach and the SCSA is particularly spectacular for all anomalous dimensions up to two-loop order, see Table~\ref{tablexpoC30} of App.~E, where we have gathered all perturbative values of $\eta$,  and Table~\ref{tablexpoC3} of App.~E. The agreement is less pronounced with the NPRG approach except for $\eta_4$, see Table~\ref{tablexpoC3bis} of App.~E. 

Due to the complexity of the numerators at three-loop order, the comparison between these three approaches is not obvious for an arbitrary codimension $d_c$. We thus consider the series in the physical case $d_c=1$. 
In this case one recovers(see \cite{coquand20a}) at the fixed point $P_2'$ the exact agreement between the perturbative computation and the SCSA up to order $\epsilon^2$, see Table~\ref{tableP2}. The comparison, at three-loop order, with the SCSA is less satisfying as  with this last method one observes a strong increase of the $\epsilon^3$ coefficient with respect to the $\epsilon^2$ one. Conversely the agreement with the NPRG approach is not very  good at order $\epsilon^2$ -- for unclear reasons one finds the wrong sign at this order (see \cite{coquand20a}) -- but satisfying at three-loop order. Nevertheless at the fixed point $P_3$ (see Table~\ref{tableP3}), and especially at $P_4$ (see Table~\ref{tableP4}), the agreement between {\it all} approaches is particularly good  -- up to three significant digits with the SCSA, a little bit less with the NPRG -- at two-loop but also at three-loop orders.

We now discuss our result for $\eta_4$ associated to the fully stable fixed point $P_4$.  The  series giving $\eta_4$ are meant to be asymptotic and resummation techniques should be used in principle. However, as far as the three-loop order is considered, the series appears to be convergent thanks to the large regular denominator structure  of the form  $\epsilon^l/(n+d_c)^{2l-1}$, with $l$ the loop order and $n=4,20,24$. We expect the singular nature of the series to show up at higher orders when  the numerators  become large enough  to  overcome the denominators. Since, up to three loop order, the coefficients are small and decreasing, asymptotic analysis allows us to truncate the series (up to the smallest coefficient) with minimum error and without any resummation. This is the co-called ``optimal truncation rule'' (see, e.g.,\cite{boyd99}), and is supposed to give a good  approximation to an asymptotic series. At some higher loop order, one should reach some increasing coefficients, and  need performing resummations to get better approximations.  At $\epsilon=2$ one gets successively at one-, two-, and three-loop orders: 
\begin{equation}
\begin{array}{ll}
&\displaystyle \eta_4^{1l}=\frac{24}{25}=0.96\, , \hspace{1cm} 
\eta_4^{2l}=\frac{2856}{3125}\simeq 0.9139\, , \nonum \\
\\
&\displaystyle \eta_4^{3l}=\frac{2856}{3125}+\frac{4 (568241 - 1286928\, \zeta(3))}{146484375}\simeq 0.8872\,. 
\end{array}
\end{equation}
Clearly, $\eta_4$ gets closer and closer, order by order, to the values obtained directly in $D=2$ from the SCSA, where $\eta_4=0.821$ and the NPRG where $\eta_4= 0.849$. Moreover, the three-loop order value we find is already compatible with the value obtained with these techniques and within  the range $[0.72,0.88]$ where  many numerical values lie \cite{guitter90,zhang93,bowick96,gompper97,los09,troster13,wei14,troster15,los16,kosmrlj17,hasik18}.

Let us finally discuss our predictions about the anomalous  and Poisson ratios $\delta$ and $\nu$. Our results, as well as those already obtained at two-loop order \cite{coquand20a}, display (small) corrections with respect to  the leading order values. They contrast with  those obtained within both the SCSA and NPRG approaches  that lead to $\delta \nu=0$ and, thus, $\nu=-{1/3}$. Several comments are required. First, we have indicated that our results for $\delta$ and $\nu$ are model, and thus very likely,  scheme-dependent. It is very desirable to get  scheme-independent values for these quantities in order to compare our results with those obtained from  other methods in a relevant way.  Then, if these  order-dependent corrections persist, one should inquire about higher-order contributions of the  series providing $\nu$. Finally, one should also inquire about the contributions that have been neglected within the SCSA and NPRG approaches, that could affect the leading order results. More generally,  this raises the question  of the deviation of the Poisson ratio with respect to the value $-1/3$, a fact that  has been recently proposed, notably  in \cite{burmistrov18b}.


 {\it Conclusion.} We have investigated the flat phase of polymerized membranes at three-loop order by means of a weak-coupling, perturbative approach of two complementary models. We have determined the RG equations, their fixed points and the associated anomalous dimensions. The agreement between the results obtained from the two models shows that we have obtained  an unambiguous control of the renormalization procedure in both models. The details of the -- involved -- computations will be given in a forthcoming publication \cite{coquand21c}. 

From our results, the order-by-order agreement found between the perturbative approach and the nonperturbative ones when the later are re-expanded in powers of $\epsilon$, which has already been observed at two-loop order, is strikingly confirmed at three-loop order. Let us add that, though our perturbative results are supposed to be valid at weak coupling, they are nevertheless exact order by order in the coupling constants. Hence, they serve as a benchmark for nonperturbative and numerical techniques that also rely on their own sets of approximations. In this context, a remarkable feature of our results with respect to the one- and two-loop  orders  calculations, is that the value found for the three-loop order critical exponent $\eta_4$ in $D=2$ (without any resummation of the $\epsilon$-series) is in {\it quantitative} agreement with the usually accepted ones from (all orders) nonperturbative methods.

Such a spectacular agreement raises the question of its very origin and, thus, of the very nature of the field theory describing the flat phase of membranes. This salient feature can be explained by the smallness of the coefficients found in the $\epsilon$-series, see Eqs.~(\ref{seta3pert}) and (\ref{seta4pert}). As can be seen from these equations, the coefficients even get smaller with increasing the loop order thus seemingly alleviating the asymptotic nature of the series, at least up to three loops. A higher order examination of these coefficients would be very interesting but is beyond the scope of this paper.


Finally, one can note that recent attempts have been made to probe more deeply the nonperturbative structure of the theory, notably concerning the relation between scale invariance and conformal symmetry, see \cite{mauri21}. The result could be considered as deceptive: the scale invariance at the infrared fixed point   not  being promoted to conformal invariance at the fixed point, the use of methods such as conformal bootstrap techniques, seems to be excluded. The crux of the matter  still lies ahead of us.


{\it Acknowledgements.} S.M.\ warmly thanks P.~Marquard and S.~Moch, the organizers of the CAPP school 2021, together with the speakers T.~Hahn and V.~Magerya, for sharing long term experience, and precious advice  on multiloop computations, including valuable Mathematica tips from which this paper greatly benefits. D.M.\ thanks O.~Coquand for discussions and for a careful reading of the manuscript. S.T.\ thanks M.~Kompaniets for useful discussions.



\appendix

\begin{widetext}

\section{Two-field model: three-loop order RG equations}

We give here the three-loop order RG equations for the -- dimensionless renormalized -- coupling constants entering action of Eq.~(\ref{action}). Note that for simplicity we omit the R indices on the renormalized coupling constants.
\begin{equation}
\begin{split}
\beta_{\mu}=& -\epsilon\mu + 2 \mu\, \eta+ \frac{d_c\,\mu^2}{6(16\pi^2)}\bigg(1+\frac{227}{180}\, \eta^{(0)}\bigg) 
+ \frac{d_c\,\mu^4}{144(96\pi^2)^3(\lambda+2\mu)^2} \times \\
&\Big(\lambda^2 \big(22839\, d_c-59616\, \zeta (3)+128672\big) + 8 \lambda \mu \big(2975\, d_c+36 (713-63\, \zeta (3))\big) + \mu^2 \big(8451\, d_c+25920\, \zeta (3)+67744\big)\Big)\\
\\
\beta_{\lambda}=&-\epsilon\lambda + 2\lambda\, \eta + \frac{d_c\big(6\lambda^2+6\lambda\mu+\mu^2\big)}{6(16\pi^2)} -\frac{d_c \big(378\lambda^2-162\lambda\mu-17\mu^2\big)}{1080\, (16\pi^2)}\, \eta^{(0)}-\frac{d_c^2\,\mu(3\lambda+\mu)^2}{36(16 \pi^2)^2} + \frac{d_c\,\mu^2}{144(96\pi^2)^3(\lambda+2\mu)^2} \times \\
&\Big(\lambda^2 \mu^2 \left(6588\, d_c^2+416667\, d_c-51840 \zeta(3)+109400\right) +54 \lambda^3 \mu \big(d_c (84\, d_c+9539)-5184 \zeta(3)+8512\big)\\
&+36 \lambda^4 \big(9 \,d_c (3\, d_c+722)-3888 \zeta(3)+7876\big)+2 \lambda \mu^3 \big(d_c (1512\,d_c+74309)+648 (52 \zeta(3)-99)\big) \\
&+ \mu^4 \big(9 \, d_c (48\, d_c+2071)+10368 \zeta(3)-52592\big)\Big)\\
\end{split}
\label{2loopslam}
\end{equation}
where the field anomalous dimension $\eta$ is given by:
\begin{equation}
\begin{split}
&\eta=\eta^{(0)}+\eta^{(1)}+\eta^{(2)}=\frac{5\mu(\lambda+\mu)}{16\, \pi^2(\lambda+2\mu)}-\frac{\mu^2\Big((340+39\,d_c)\lambda^2+4(35+39\,d_c)\lambda\mu+(81\,d_c-20)\mu^2\Big)}{72\, (16 \pi^2)^2(\lambda+2\mu)^2}+\frac{\mu^3}{24(96\pi^2)^3(\lambda+2\mu)^3} \times \\
& \hspace{-0.7cm}\Big(\lambda^2 \mu \left(-19020\, d_c^2-1296(106\, d_c+179) \zeta(3)+210129\, d_c+264058\right)+ \lambda^3 \left(-12745\, d_c^2-1296 (28\, d_c+25) \zeta(3)+97175\, d_c+135430\right)\\
&\hspace{-0.6cm} + \lambda \mu^2 \left(-8640 \,d_c^2-2592 (43\, d_c+158)\zeta(3)+122337\, d_c+350372\right) +\mu^3 (-5184 (2 \,d_c+45)\zeta (3)+5(199-307\, d_c)\, d_c+211700)\Big)
\label{eta}
\end{split}
\end{equation}

\section{Two-field model: fixed points coordinates and anomalous dimensions}

We give here the coordinates of the fixed points $P_3$ and $P_4$ and the associated anomalous dimensions $\eta_3$ and $\eta_4$ obtained at three-loop order with the two-field model. 

\begin{table}[h!]
\begin{center}
\begin{tabular}{|c|l|}
\hline \vcrule 
$\mu^*_3$ & $\displaystyle\frac{96\pi^2\, \epsilon}{20+d_c} +\frac{80\pi^2(-d_c+232)}{3(20+d_c)^3}\, \epsilon^2$ \\ \vcrule 
& \hspace{1.1cm} $\displaystyle + \frac{2 \pi ^2 \left(1985 \, d_c^3+787720\, d_c^2+12700408\, d_c-32499040-2592 (d_c+20) (223 \, d_c-1018)\zeta(3)\right)}{27 (20+d_c)^5}\, \epsilon^3$ \\ 
\hline \vcrule 
$\lambda^*_3$ & $\displaystyle\ -\frac{48\pi^2 \,\epsilon}{20+d_c}-\frac{8\pi^2(9\, d_c^2+265\, d_c+2960)}{3(20+d_c)^3}\, \epsilon^2 $ \\ \vcrule 
& \hspace{1.4cm} $ \displaystyle +\frac{2 \pi ^2 \left(-d_c (d_c(d_c (243\, d_c+15649)+ 
737840)+9844004)+5593520+1296 \,(d_c+20) (223 \,d_c-1018)\zeta(3)\right)}{27 (20+d_c)^5}\, \epsilon ^3$ \\ 
\hline \vcrule 
$\eta_3$ & $\displaystyle\ \frac{10\,\epsilon}{20+d_c}-\frac{d_c(37\,d_c+950)}{6(20+d_c)^3}\,\epsilon^2$ \\ \vcrule 
& \hspace{1.2cm} $ \displaystyle + \frac{\, d_c\left(-465 \, d_c ^3-71410 \, d_c^2-1120214 \, d_c+1381320 +1296 (d_c+20) (32 \, d_c-273)\zeta(3)\right)}{216 (20+d_c)^5}\, \epsilon ^3 $ \\
\hline
\end{tabular}
\end{center}
\vspace{-0.25cm}
\caption{Coordinates $\mu^*_3$ and $\lambda^*_3$ of the fixed point $P_3$ and the corresponding field anomalous dimension $\eta_3$ at order $\epsilon^3$ obtained from the two-field model (this work).}
\label{tablexpo3}
\end{table}

\begin{table}[h!]
\begin{center}
\begin{tabular}{|c|l|}
\hline \vcrule
$\mu^*_4$ & $\displaystyle\frac{96\pi^2 \, \epsilon}{24+d_c}-\frac{32\pi^2(47 d_c+228)}{5(24+d_c)^3}\, \epsilon^2 $\\ \vcrule 
& $\displaystyle \hspace{1.2cm} -\frac{12 \pi ^2 \left(2305 \, d_c^3-307264 \, d_c^2-5977136 \, d_c+58197120+4608 (d_c+24) (64 \, d_c-465) \zeta(3)\right)}{125 (24+d_c)^5}\, \epsilon ^3 $ \\ 
\hline \vcrule 
$ \lambda^*_4$ & $\displaystyle\ -\frac{32\pi^2 \,\epsilon}{24+d_c}+\frac{32\pi^2(19\, d_c+156)}{5(24+d_c)^3}\, \epsilon^2 $ \\ \vcrule 
& $\displaystyle \hspace{1.5cm} +\frac{4 \pi ^2 \left(1865 \, d_c^3-1054052 \, d_c^2-15313488 \, d_c+243792000+300672 (d_c+24) (3 \, d_c-20) \zeta(3)\right)}{375 (24+d_c)^5}\, \epsilon ^3$ \\ 
\hline \vcrule 
$\eta_4$ & $\displaystyle\ \frac{12\,\epsilon}{24+d_c}-\frac{6 d_c(d_c+29)}{(24+d_c)^3}\,\epsilon^2 $ \\ \vcrule
& $\displaystyle \hspace{1.3cm} + \frac{\, d_c \left(-2000 \, d_c^3-231029 \, d_c^2-3531426 \, d_c+17970480+1296 (d_c+24) (67 \, d_c-1060) \zeta(3)\right)}{750 (24+d_c)^5}\, \epsilon ^3$ \\
\hline
\end{tabular}
\end{center}
\vspace{-0.25cm}
\caption{Coordinates $\mu^*_4$ and $\lambda^*_4$ of the stable fixed point $P_4$ and the corresponding field anomalous dimension $\eta_4$ at order $\epsilon^3$ obtained from the two-field model (this work).}
\label{tablexpo4}
\end{table}

\newpage

\section{Flexural effective model: three-loop order RG equations}

We give here the three-loop order RG equations for the -- dimensionless renormalized -- coupling constants entering action of Eq.~(\ref{actioneff}). Again we omit the R indices on the renormalized coupling constants.

\begin{equation}
\begin{split}
&\beta_{\mu}=-\epsilon\mu + 2 \mu\, \eta+\frac{d_c\,\mu^2}{6 (16\pi^2)}\bigg(1+\frac{107\, b +574\,\mu}{216\, (16 \pi^2)}\bigg) +\frac{d_c\, \mu^2}{5184(96\pi^2)^3} \times \\
& \Big(b^2 \big(34987 \, d_c-384 \big(459 \zeta(3)-296)\big)+32\, b\,\mu \big(980\, d_c+2754\zeta (3)+3801\big)+4\mu^2 \big(5317 \,d_c+80352 \zeta(3)+93600\big)\Big)
\\
\\
&\beta_b=-\epsilon b + 2 b\, \eta\ +\frac{5 d_c\, b^2}{12(16\pi^2)}\bigg(1+\frac{178\, \mu-91 b}{216\, (16 \pi^2)}\bigg) + \frac{d_c\, b^2}{10368(96\pi^2)^3} \times \\
& \Big(b^2 \big(371495 \, d_c-228096 \zeta(3)+614832\big)+2240 \, b \, \mu \big(7 \, d_c-6 (54 \zeta(3)+5)\big)+4 \mu^2 \big(87893 \, d_c+425088 \zeta(3)-248616\big)\Big)
\end{split}
\label{2loopseff}
\end{equation}
and: 
\begin{equation}
\begin{split}
\eta&=\frac{5(b+2\mu)}{6(16\pi^2)} + \frac{5\,(15\,d_c - 212)\,b^2 + 1160\,b\,\mu - 4\,(111\,d_c - 20)\,\mu^2}{2592 (16\pi^2)^2} +\frac{1}{5184(96\pi^2)^3} \times \\
& \Big(b^3 \left(-41625 \, d_c^2+5184 (9\, d_c+16)\zeta(3)+180563 \, d_c+516252\right)-6\, b^2\, \mu \big(18144 (5 \, d_c+2)\zeta(3)-56445 \, d_c+221204\big)+ \\
&12\, b\, \mu^2 \big(-1296 (50 \, d_c+57)\zeta(3)+82681 \, d_c+108974\big)-8 \mu^3 \left(1395 \, d_c^2-1296 (96 \, d_c-509) \zeta(3)+188605 \, d_c-652398\right)\Big)
\label{etaeff}
\end{split}
\end{equation}

\newpage

\section{Flexural effective model: fixed points coordinates and anomalous dimensions}

We give here the coordinates of the fixed points $P_2'$, $P_3$ and $P_4$ and the associated field anomalous dimensions obtained at three-loop order with the flexural effective model. 

\begin{table}[h!]
\begin{center}
\begin{tabular}{|c|l|}
\hline \smallvcrule
${\mu}'^{*}_2$ & \hspace{6.5cm} $0$ \\ 
\hline \vcrule 
${b}'^*_2$ & $\displaystyle\frac{192\pi^2 \, \epsilon}{5(4+d_c)}+\frac{32\pi^2(61\, d_c+424)}{75(4+d_c)^3}\, \epsilon^2 $ \\ \vcrule 
& $\displaystyle \hspace{1.4cm} +\frac{4 \pi ^2 \left(-204995 \, d_c^3-2008224 \, d_c^2-5344224 \, d_c-1068992+41472 (d_c-8) (d_c+4) \zeta(3)\right)}{16875 (4+d_c)^5}\, \epsilon ^3$ \\
\hline \vcrule 
$\eta_2'$ & $\displaystyle \frac{2\,\epsilon}{4+d_c}+\frac{d_c(d_c-2)}{6(4+d_c)^3}\,\epsilon^2 $ \\ \vcrule
& $\displaystyle \hspace{1cm} +\frac{d_c \big(-d_c (d_c (13875 \, d_c+113044)+257236)-216240+5184 (d_c+4) (3 \, d_c+20)\zeta(3)\big)}{27000 (4+d_c)^5}\, \epsilon ^3$ \\
\hline
\end{tabular}
\end{center}
\vspace{-0.25cm}
\caption{Coordinates ${\mu}'^{*}_2$ and ${b}'^*_2$ of the fixed point $P_2'$ and the corresponding field anomalous dimension $\eta_2'$ at order $\epsilon^3$ obtained from the flexural effective model (this work).}
\label{tablexpoeff2}
\end{table}

\begin{table}[h!]
\begin{center}
\begin{tabular}{|c|l|}
\hline \vcrule
$\mu^*_3$ & $\displaystyle\frac{96\pi^2 \, \epsilon}{20+d_c}-\frac{80\pi^2(13\, d_c+8)}{3(20+d_c)^3}\, \epsilon^2 $ \\ \vcrule
& $\displaystyle \hspace{1.2cm} +\frac{2 \pi ^2 \left(263 \, d_c^3+699880 \, d_c^2+10648408\, d_c-52179040-2592 (d_c+20) (223 \, d_c-1018) \zeta(3)\right)}{27 (20+d_c)^5}\, \epsilon ^3$ \\ 
\hline \smallvcrule
$b^*_3$ & \hspace{6.5cm} $0$ \\
\hline \vcrule
$\eta_3$ & $\displaystyle \frac{10\,\epsilon}{20+d_c}-\frac{d_c(37\, d_c+950)}{6(20+d_c)^3}\,\epsilon^2 $\\ \vcrule
& $\displaystyle \hspace{1.2cm} +\frac{\, d_c \left(-465 \, d_c^3-71410 \, d_c^2-1120214 \, d_c+1381320+1296 (d_c+20) (32 \, d_c-273) \zeta(3)\right)}{216 (20+d_c)^5}\, \epsilon ^3$ \\ 
\hline
\end{tabular}
\end{center}
\vspace{-0.25cm}
\caption{Coordinates $\mu^*_3$ and $b^*_3$ of the fixed point $P_3$ and the corresponding field anomalous dimension $\eta_3$ at order $\epsilon^3$ obtained from the flexural effective model (this work).}
\label{tablexpoeff3}
\end{table}

\begin{table}[h!]
\begin{center}
\begin{tabular}{|c|l|}
\hline \vcrule
$\mu^*_4$ & $\displaystyle\frac{96\pi^2 \, \epsilon}{24+d_c}-\frac{32\pi^2(77\,d_c+948)}{5(24+d_c)^3}\, \epsilon^2 $ \\ \vcrule
& $\displaystyle \hspace{1.2cm} + \frac{16 \pi ^2 \left(-5105 \, d_c^3+734094 \, d_c^2+14820156 \, d_c-121525920-10368 (d_c+24) (64 \, d_c-465) \zeta(3)\right)}{375 (24+d_c)^5}\, \epsilon ^3$ \\ 
\hline \vcrule
$b^*_4$ & $\displaystyle \frac{192\pi^2 \,\epsilon}{5(24+d_c)}+\frac{64\pi^2(121 d_c+3804)}{25(24+d_c)^3}\, \epsilon^2 $ \\ \vcrule
& $\displaystyle \hspace{1.6cm} +\frac{16 \pi ^2 \left(-48445 \, d_c^3-536634 \, d_c^2+6606936 \, d_c-169636032-5184 (d_c+24) (247 \, d_c-2076) \zeta(3)\right)}{1875 (24+d_c)^5}\, \epsilon ^3$ \\
\hline \vcrule
$\eta_4$ & $\displaystyle \frac{12\,\epsilon}{24+d_c}-\frac{6 d_c(d_c+29)}{(24+d_c)^3}\,\epsilon^2 $ \\ \vcrule
& $\displaystyle \hspace{1.2cm} +\frac{\, d_c \left(-2000 \, d_c^3-231029 \, d_c^2-3531426 \, d_c+17970480+1296 (d_c+24) (67 \, d_c-1060) \zeta(3)\right)}{750 (24+d_c)^5}\, \epsilon ^3$\\ 
\hline
\end{tabular}
\end{center}
\vspace{-0.25cm}
\caption{Coordinates $\mu^*_4$ and $b^*_4$ of the stable fixed point $P_4$ and the corresponding field anomalous dimension $\eta_4$ at order $\epsilon^3$ obtained from the flexural effective model (this work).}
\label{tablexpoeff4}
\end{table}

\section{Three-loop, SCSA and NPRG: the anomalous dimensions}

We provide below the anomalous dimensions obtained at various fixed points by means of the three-loop order, SCSA and NPRG approaches, respectively. 

\begin{table}[htbp]
\begin{center}
\begin{tabular}{|c|l|}
\hline \smallvcrule
& \multicolumn{1}{c|}{Three-loop} \\ 
\hline \hline \vcrule
$\eta_2'$ & $\displaystyle \frac{2\,\epsilon}{4+d_c}+\frac{d_c(d_c-2)}{6(4+d_c)^3}\,\epsilon^2 $ \\ \vcrule
& $\displaystyle \hspace{1cm} +\frac{d_c \big(-d_c (d_c (13875 \, d_c+113044)+257236)-216240+5184 (d_c+4) (3 \, d_c+20)\zeta(3)\big)}{27000 (4+d_c)^5}\, \epsilon ^3$ \\ 
\hline \vcrule
$\eta_3$ & $\displaystyle \frac{10\,\epsilon}{20+d_c}-\frac{d_c(37 d_c+950)}{6(20+d_c)^3}\,\epsilon^2 $\\ \vcrule
& $\displaystyle \hspace{1.2cm} +\frac{\, d_c \left(-465 \, d_c^3-71410 \, d_c^2-1120214 \, d_c+1381320+1296 (d_c+20) (32 \, d_c-273) \zeta(3)\right)}{216 (20+d_c)^5}\, \epsilon ^3$ \\
\hline \vcrule 
$\eta_4$ & $\displaystyle \frac{12\,\epsilon}{24+d_c}-\frac{6 d_c(d_c+29)}{(24+d_c)^3}\,\epsilon^2 $ \\ \vcrule
& $\displaystyle \hspace{1.2cm} +\frac{\, d_c \left(-2000 \, d_c^3-231029 \, d_c^2-3531426 \, d_c+17970480+1296 (d_c+24) (67 \, d_c-1060) \zeta(3)\right)}{750 (24+d_c)^5}\, \epsilon ^3$ \\ 
\hline
\end{tabular}
\end{center}
\vspace{-0.25cm}
\caption{field anomalous dimensions $\eta_2'$, $\eta_3$ and $\eta_4$ at order $\epsilon^3$ obtained from the three-loop order approach (this work).}
\label{tablexpoC30}
\end{table}

\begin{table}[htbp]
\begin{center}
\begin{tabular}{|c|l|}
\hline \smallvcrule
& \multicolumn{1}{c|}{SCSA} \\
\hline \hline \vcrule
$\eta_2'$ & $ \displaystyle \frac{2\,\epsilon}{4+d_c}+\frac{d_c(d_c-2)}{6(4+d_c)^3}\,\epsilon^2 
-\frac{d_c\left(37 \, d_c^3+186
\, d_c^2-120 \, d_c-800\right)}{72
(4+d_c)^5} \epsilon ^3$ \\
\hline \vcrule
$\eta_3$ & $\displaystyle \frac{10\,\epsilon}{20+d_c}-\frac{d_c (37d_c+890)}{6(20+d_c)^3}\,\epsilon^2-\frac{d_c\left(155 \, d_c^3+5706
\, d_c^2+138840 \, d_c+1914400\right)}{72
(20+d_c)^5} \epsilon ^3$ \\
\hline \vcrule
$\eta_4$ & $\displaystyle \frac{12\,\epsilon}{24+d_c}-\frac{6 d_c (d_c+30)}{(24+d_c)^3}\,\epsilon^2-\frac{d_c \left(8 \, d_c^3+273
\, d_c^2+5328 \, d_c+96768\right)}{3 (24+d_c)^5} \epsilon ^3$ \\
\hline
\end{tabular}
\end{center}
\vspace{-0.25cm}
\caption{field anomalous dimensions $\eta_2'$, $\eta_3$ and $\eta_4$ at order $\epsilon^3$ obtained from the SCSA \cite{ledoussal92,ledoussal18}.}
\label{tablexpoC3}
\end{table}

\begin{table}[htbp]
\begin{center}
\begin{tabular}{|c|l|}
\hline \smallvcrule
& \multicolumn{1}{c|}{NPRG} \\ 
\hline \hline \vcrule 
$\eta_2'$ & $\displaystyle {2\eps \over 4+d_c}+ {d_c(10+3\,d_c)\over 12(4+d_c)^3}\eps^2 + \frac{d_c \left(9 \, d_c^3+79
\, d_c^2+234 \, d_c+232\right) }{144 (4+d_c)^5}\epsilon^3$ \\
\hline \vcrule
$\eta_3$ & $\displaystyle \frac{10\,\epsilon}{20+d_c}-\frac{d_c(69\,d_c+1430)}{12(20+d_c)^3}\epsilon^2 +\frac{d_c\left(63 \, d_c^3-5177
\, d_c^2-303630 \, d_c-3507800 \right)}{144 (20+d_c)^5}\epsilon ^3$ \\ 
\hline \vcrule
$\eta_4$ & $\displaystyle\frac{12\,\epsilon}{24+d_c} -\frac{\,d_c(11\,d_c+276)}{2(24+d_c)^3}\,\epsilon^2 +\frac{d_c\left(2 \, d_c^3-81 \, d_c^2-8148 \, d_c-121824\right)}{4(24+d_c)^5} \epsilon ^3$ \\ 
\hline
\end{tabular}
\end{center}
\vspace{-0.25cm}
\caption{field anomalous dimensions $\eta_2'$, $\eta_3$ and $\eta_4$ at order $\epsilon^3$ obtained from the NPRG \cite{kownacki09}.}
\label{tablexpoC3bis}
\end{table}

\end{widetext}

\end{document}